\documentclass[mathleft
]{an}
\usepackage{graphicx}
\usepackage{epstopdf}
\usepackage{times}
\begin{document}

\Pagespan{206}{209}
\Yearpublication{2009}%
\Yearsubmission{2009}%
\Month{11}%
\Volume{330}%
\Issue{2/3}%
\DOI{10.1002/asna.200811157}%

\title{Identifying Compact Symmetric Objects from the VLBA Imaging and Polarization Survey}

\author{S.E. Tremblay\inst{1}\fnmsep\thanks{Corresponding author:
  \email{tremblay@unm.edu}\newline}
\and  G.B. Taylor\inst{1,2}
\and  J.F. Helmboldt\inst{3}
\and C.D. Fassnacht\inst{4}
\and R.W. Romani\inst{5}
}
\titlerunning{Identifying CSOs from VIPS}
\authorrunning{S.E. Tremblay et al.}
\institute{
Department of Physics and Astronomy, University of New Mexico, Albuquerque, NM 87131
\and 
National Radio Astronomy Observatory, Socorro,  NM 87801
\and 
Naval Research Laboratory, Code 7213, Washington, DC 20375
\and
Department of Physics, University of California, Davis, CA 95616
\and 
Department of Physics, Stanford University, Stanford, CA 94305}

\received{2008 Dec 8}
\accepted{2008 Dec 18}
\publonline{2009 Feb 15}


\abstract{%
  Compact Symmetric Objects (CSOs) are small (less than 1 kpc) radio sources which have symmetric double lobes or jets. The dominant theory for the small size of these objects is that they are young radio sources which could grow into larger radio galaxies, but the currently small population of known CSOs makes it difficult to definitively determine whether or not this is the case. While a greater number of Gigahertz peaked sources can be identified by sifting through spectral surveys, this yields none of the dynamics of the sources, and also brings Quasars into the sample, which although interesting are peaked around 1 Gigahertz for very different reasons. We have used the 5 GHz VLBA Imaging and Polarization Survey (VIPS) to identify 103 CSO candidates morphologically, and are following up on these sources with multifrequency VLBA observations to confirm CSO identifications and to study their dynamics. The identification of candidates from within the survey will be discussed, as well as preliminary results from the follow-up observations.}

\maketitle

\section{Introduction}
 
CSOs are a class of small ($<1$ kpc) radio sources that have lobe emission on both sides of an active core, which are oriented at large angles to the line of sight (Wilkinson et al. 1994).  CSOs are currently thought to be very young ($<10^4$ yr) objects (Readhead et al. 1996b) which may evolve into Fanaroff-Riley II radio galaxies (Fanaroff \& Riley 1974; Readhead et al. 1996a). Since the jets are at such a large angle with the line of sight, there is often little to no Doppler boosting of the approaching jet (Wilkinson et al. 1994), which means that the receding jet can contain up to half of the observed flux density which makes it possible to detect an obscuring torus (or circumnuclear disk) such as is seen in J1945+7055 (Peck \& Taylor 2001). The random orientation angle of the jets makes it possible to evaluate if these disks that are predicted by unified schemes of AGN (Antonucci 1993) meet the necessary criteria.  Furthermore, the age of CSOs can be measured from the rate at which each hotspot (the working surface of the jet) advances into the external medium.  The ability to determine a kinematic age, as well as using other techniques such as computing an age based on their spectra, makes CSOs a good sample with which to study the evolution of AGN.

Some CSOs, such as 0108+388 (Taylor et al. 1995), have extended emission connecting the larger components, which, given the velocities measured in these objects, indicates either continued activity over millions of years, or at least sporadic activity on these timescales. By assembling a larger sample of CSOs with uniform low frequency VLA measurements, we will constrain the duty cycle of radio activity.  Using the 5 GHz VIPS and the 5 GHz follow-up observations, and multiple epochs when available (e.g., for CJF sources), we will estimate kinematic ages for those CSOs with moving hot spots.  This will inform us about the age distribution of CSOs.  Preliminary studies based on a small number of CSOs with measured ages reveal an abundance of very young sources, indicating that most CSOs cease activity after only a few hundred years (Gugliucci et al. 2005).

So far, there have been only a few detections of significantly polarized CSOs (Gugliucci et al. 2005). Obtaining Faraday rotation measures of these and newly detected polarized CSOs would be an excellent tool for probing the environment surrounding these young AGN.

To date the closest compact Supermassive Binary Black Hole  (SBBH) system, 0402+379, has been found in a CSO (Rodriguez et al. 2006).   These systems are of great interest as potential gravitational wave emitters and as a laboratory for testing strong gravity.  They may also describe an essential sequence in the evolution of galaxies as the most practical way to grow supermassive black holes is via mergers. Further studies of CSOs are warranted if only to discover more SBBH systems. 

Currently, there have been a small number of CSOs discovered ($<50$) and studied using the resolution provided by VLBI imaging. Furthermore, those that have been studied are from a variety of samples. This has made it difficult to study CSOs as a whole, instead of the objects' individual characteristics. The recent VLBA Imaging and Polarization Survey (VIPS; Taylor et al. 2005; Helmboldt et al. 2007) has identified a large number of CSO candidates (103) out of the largest imaging survey yet undertaken with VLBI  (1127 sources) and will provide an excellent basis to look for and study CSOs in detail. It is hoped that with proposed follow-up observations we can double or triple the number of confirmed CSOs, giving a better sample with which to discuss their nature. This large sample should also contain a range in ages of the objects, allowing us to study their evolution from 10s to 1000s of years.  Furthermore, the VIPS survey has been conducted in full polarization so that more polarized CSOs should be identified, and the completeness of the survey will make it possible to describe statistical statements about CSOs in general.

\section{The Sample}

\subsection{VIPS}
An overview of VIPS will be presented here, while the details can be found in Helmboldt et al. (2007). 
 
\subsubsection{VIPS Sample}
The VIPS sample was defined by creating a sub-sample of the Cosmic Lens All-Sky Survey (CLASS; Myers et al. 
2003) sample, which consists of over 12,000 flat-spectrum objects ($\alpha > -0.5$ between 4.85 GHz and a lower frequency where $F_\nu \propto \nu^\alpha$). The sample was then restricted to lie within the fifth Data Release of the Sloan Digital Sky Survey (SDSS; Adelman-McCarthy et al. 2007), which \linebreak makes a "footprint" that covers approximately 8000 square degrees on the sky. Additionally, the declination of sources was required to be greater than $15^\circ$ since good \emph{u-v} coverage is quite difficult for sources lower on the sky than this using the VLBA.

\subsubsection{Observations}

VIPS sources were observed in 18 different 11-hour observing runs between January and August in 2006. Each source was observed over 10 separate scans to maximize \emph{u-v} coverage with a total integration time around 500 s. Observations were conducted with four 8 MHz wide IFs with full polarization. 

\subsubsection{Calibration and Imaging}

Both the calibration and image mapping processes utilized scripts to automate them as much as possible. Initial data flagging and calibration were performed in AIPS (Greisen 2003) using the VLBA data calibration pipeline (Sjouwerman et al. 2005). The imaging was carried out in DIFMAP (Shepherd 1997) using similar scripts to those described in Taylor et al. (2005). The automated imaging only 'failed' in a few cases ($<1\%$) and these sources were re-imaged manually in DIFMAP.

\subsubsection{Source Classification}

Once the sources were all imaged, morphological classification was the next step in the analysis. Once again the process was automated to accommodate the large number of sources in the study. Multiple component Gaussian models were first created for all of the detected sources in AIPS using the SAD task. The sources were then individually classified based on the following criteria:

\begin{itemize}
\item If the source can be modeled with a single Gaussian component that contains at least 95\% of the total flux density then the source is flagged as a single-component source.
\item If a non-single-component source's two brightest Gaussian components contain at least 95\% of the total flux density then the source is flagged as a double source.
\item Sources not flagged as either single or double sources are then subsequently flagged as multiple component sources. 
\end {itemize}

\noindent Sources are further classified from these rough categories to their final classications.
\begin{itemize}
\item Single-component sources which have an axis ratio less than 0.6 are classified as core-jets, otherwise they are classified as point sources (PSs).
\item Double sources in which the flux density of the two primary components agree to within a factor of 2.5 are classified as CSO candidates (CSOs).
\item If the dominant components (the components containing 95\% of the total flux density of the source) of a multiple component source line up then the source is flagged as a core-jet, otherwise the source is complex (CPLX).
\item Core-jets longer than 6 mas are classified as long jets (LJETs), otherwise they are placed into the category of short jets (SJETs).
\item If a LJET source is longer than 12 mas and its brightest Gaussian component is within 3 mas of the center of the structure, then the source is reclassified as a CSO candidate.
\end{itemize}

\subsection{VIPS CSO Candidates}

As described above, there are two separate ways a source can be classified as a CSO candidate. One of these is sensitive to the 'classic double' CSOs, where the two hotspots are the brightest components in the image and are linearly oriented away from the core, which itself might or might not be detected. The other is sensitive to CSOs exhibiting continuous structure, which might be either linear or in an 'S-symmetry'.

\section{CSO Confirmation}

To judge whether or not these CSO candidates are actual CSOs, we are primarily using two criteria. These are: (1) A compact flat-spectrum object centrally located in the source. If we identify such a core in the middle of the source, it is likely a CSO. (2) We also look for symmetric radial motion from the hotspots, which normally expand out from the core at $\sim0.3$c. To look for these characteristics, we have performed multi-frequency follow-up observations (5, 8, and 15 GHz) and utilize existing archival observations of these sources. The simultaneous multi-frequency observations are crucial, since cores and hotspots each have distinctive spectral signatures.

\section{Preliminary Results}
As briefly mentioned above, we performed follow-up observations on all the VIPS CSO candidates. These were VLBA observations that took place between September 2006 and June 2008.
The following are a couple of preliminary examples from these follow-up observations. Throughout this section, we assume  H$_{0}$=73 km s$^{-1}$Mpc$^{-1}$, $\Omega_m$ = 0.27, $\Omega_\Lambda$ = 0.73. \\

\begin{figure}
 {\includegraphics[width=82 mm] {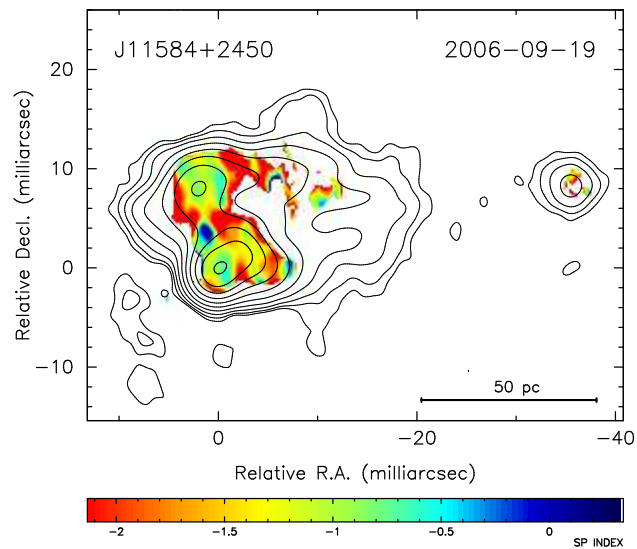}
 \caption{Multi-frequency observations of a newly identified CSO (J11584+2450). 5 GHz 
contours overlaid on a 8 to 15 GHz spectral index image, where the scale bar in the lower right is calculated using a redshift of $0.20130\pm0.00040$ (Zensus et al. 2002). Notice the flat spectrum core, as well as the symmetric dual-lobed structure in the source. Also, the emission abruptly bends 
to the west. This sudden path change, and the steep spectrum compact knot at the western 
edge is not clearly understood. (Figure adapted from Tremblay et al. 2008)}
 \label{J11584}}
 \end{figure}

\begin{figure}
 {\includegraphics[width=82 mm] {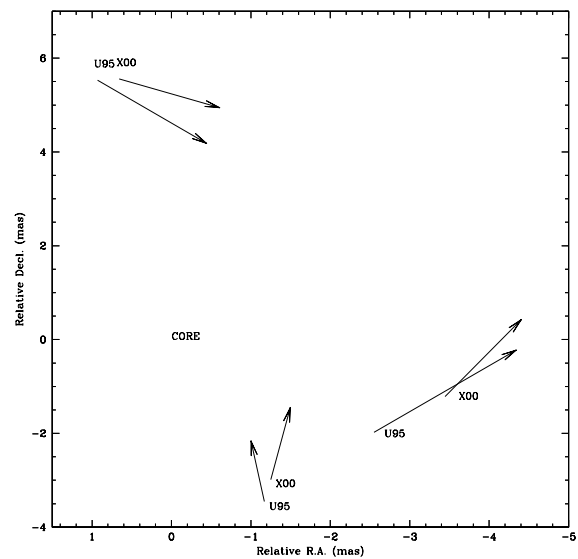}
 \caption{Relative velocity of model components. Velocity of each Gaussian model component 
is plotted (1mas = 0.2c) with the tail of each vector originating at the model components 
position at its earliest observation (1995 for 15 GHz and 2000 for 8 GHz). The model fits are 
in agreement with the contour overlay plots in showing this source to be shrinking. (Figure from Tremblay et al. 2008)}
 \label{motions}}
 \end{figure}

J11584+2450 was previously observed in multiple\linebreak epochs as part of the 2 cm Survey (Kellermann et al. 2004) and was subsequently classified as a one sided core-jet.\linebreak Multi-frequency observations in 2006 (Tremblay et al. \linebreak2008) reveal this source to actually be a CSO (Fig. \ref{J11584} shows the compact flat-spectrum core in between two steeper spectrum lobes). Even more interestingly, kinematic analysis\linebreak from five different epochs (Fig. \ref{motions}) yield an apparent contraction in this source (radial motion towards the core at $\sim0.3$c), as well as western motion (at $\sim0.2$c). If this contraction is real, and not just an observational effect, one possible interpretation could have interesting effects on our general understanding of CSOs. If there is a relative motion between the source and its clumpy ambient medium, higher external pressure would be placed on the jets if the source moves into a denser part of the medium. This could leave the jets under-pressured, causing the hotspots to temporarily contract back towards the core. If this is indeed what is happening in J11584+2450, then it implies that frustration contributes to the small sizes of at least some CSOs. 

\begin{figure}
 {\includegraphics[width=82 mm] {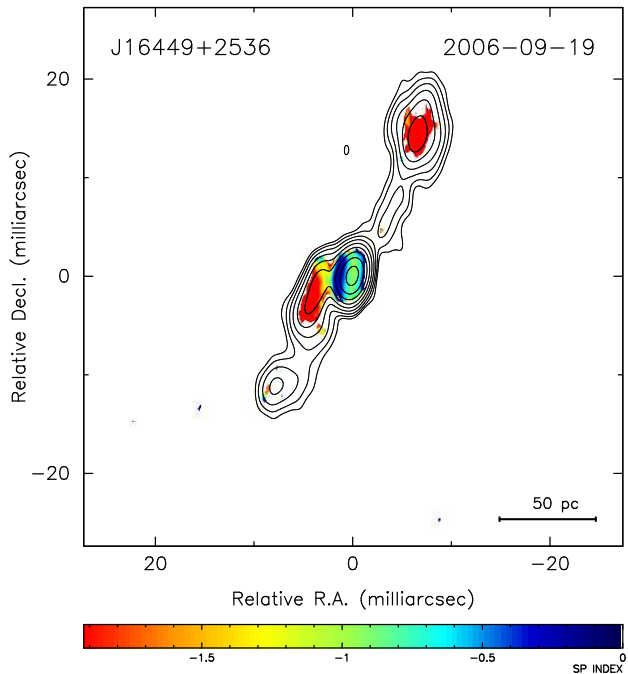}
 \caption{Multi-frequency observations of a newly identified CSO (J16449+2536). 5 GHz 
contours overlaid on a 8 to 15 GHz spectral index image, where the scale bar in the lower right is calculated using the SDSS template fit photometric redshift of $0.199408\pm0.051544$ (Csabai et al. 2003). The flat spectrum core resides between dual jets which have the 'S-symmetry' sometimes observed in CSOs.}
 \label{J16449}}
 \end{figure}

A preliminary spectral index map of the VIPS source J16449+2536 (Fig. \ref{J16449}) made from multi-frequency follow-up observations show emission oriented in  an 'S-symmetry' with a flat-spectrum core located in its center.

 \section{Future Work}
 Once all of the CSOs in VIPS have been positively identified, the complete list of VIPS CSOs will be published. Subsequent study of the properties and statistical analysis of the VIPS CSOs will also be made available once it has been completed. These observations and analysis will hopefully shed further light on CSOs, and in particular where they fit in the process of galactic evolution.

 \acknowledgements

The National Radio Astronomy Observatory is a facility of the National Science Foundation operated under cooperative agreement by Associated Universities, Inc.

\newpage

\end{document}